\documentstyle[11pt]{article}
\begin{document}
\pagestyle{plain}

%
\newcommand{\bear}{\begin{eqnarray}}
\newcommand{\ear}{\end{eqnarray}\noindent}
\newcommand{\no}{\noindent}
\date{}
\renewcommand{\theequation}{\arabic{section}.\arabic{equation}}
\renewcommand{\arraystretch}{2.5}
\newcommand{\GeV}{\mbox{GeV}}
\newcommand{\cL}{\cal L}
\newcommand{\D}{\cal D}
\newcommand{\Dhalf}{{D\over 2}}
\newcommand{\Det}{{\rm Det}}
\newcommand{\PP}{\cal P}
\newcommand{\G}{{\cal G}}
\def\R{1\!\!{\rm R}}
\def\Eins{\mathord{1\hskip -1.5pt
\vrule width .5pt height 7.75pt depth -.2pt \hskip -1.2pt
\vrule width 2.5pt height .3pt depth -.05pt \hskip 1.5pt}}
\newcommand{\symb}{\mbox{symb}}
\renewcommand{\arraystretch}{2.5}
\newcommand{\slD}{\raise.15ex\hbox{$/$}\kern-.57em\hbox{$D$}}
\newcommand{\slpartial}{\raise.15ex\hbox{$/$}\kern-.57em\hbox{$\partial$}}
\newcommand{\slG}{{{\dot G}\!\!\!\! \raise.15ex\hbox {/}}}
\newcommand{\Gd}{{\dot G}}
\newcommand{\Gund}{{\underline{\dot G}}}

\def\non{\nonumber}
\def\beqn*{\begin{eqnarray*}}
\def\eqn*{\end{eqnarray*}}
\def\sy{\scriptscriptstyle}
\def\footstrut{\baselineskip 12pt}
\def\square{\kern1pt\vbox{\hrule height 1.2pt\hbox{\vrule width 1.2pt
   \hskip 3pt\vbox{\vskip 6pt}\hskip 3pt\vrule width 0.6pt}
   \hrule height 0.6pt}\kern1pt}
\def\slash#1{#1\!\!\!\raise.15ex\hbox {/}}
\def\dint#1{\int\!\!\!\!\!\int\limits_{\!\!#1}}
\def\bra#1{\langle #1 |}
\def\ket#1{| #1 \rangle}
\def\vev#1{\langle #1 \rangle}
\def\rightvac{\mid 0\rangle}
\def\leftvac{\langle 0\mid}
\def\dps{\displaystyle}
\def\sy{\scriptscriptstyle}
\def\half{{1\over 2}}
\def\third{{1\over3}}
\def\fourth{{1\over4}}
\def\fifth{{1\over5}}
\def\sixth{{1\over6}}
\def\seventh{{1\over7}}
\def\eigth{{1\over8}}
\def\ninth{{1\over9}}
\def\tenth{{1\over10}}
\def\pa{\partial}
\def\ddtau{{d\over d\tau}}
\def\ge{\hbox{\textfont1=\tame $\gamma_1$}}
\def\gz{\hbox{\textfont1=\tame $\gamma_2$}}
\def\gd{\hbox{\textfont1=\tame $\gamma_3$}}
\def\go{\hbox{\textfont1=\tamt $\gamma_1$}}
\def\gt{\hbox{\textfont1=\tamt $\gamma_2$}}
\def\gth{\hbox{\textfont1=\tamt $\gamma_3$}} 
\def\gf{\hbox{$\gamma_5\;$}}
\def\ie{\hbox{$\textstyle{\int_1}$}}
\def\iz{\hbox{$\textstyle{\int_2}$}}
\def\id{\hbox{$\textstyle{\int_3}$}}
\def\ldop{\hbox{$\lbrace\mskip -4.5mu\mid$}}
\def\rdop{\hbox{$\mid\mskip -4.3mu\rbrace$}}
\def\eps{\epsilon}
\def\epshalf{{\epsilon\over 2}}
\def\e{\mbox{e}}
\def\g{\mbox{g}}
\def\pa{\partial}
\def\kinb{{1\over 4}\dot x^2}
\def\kinf{{1\over 2}\psi\dot\psi}
\def\expk{{\rm exp}\biggl[\,\sum_{i<j=1}^4 G_{Bij}k_i\cdot k_j\biggr]}
\def\expp{{\rm exp}\biggl[\,\sum_{i<j=1}^4 G_{Bij}p_i\cdot p_j\biggr]}
\def\expshort{{\e}^{\half G_{Bij}k_i\cdot k_j}}
\def\expabb{{\e}^{(\cdot )}}
\def\epseps#1#2{\varepsilon_{#1}\cdot \varepsilon_{#2}}
\def\epsk#1#2{\varepsilon_{#1}\cdot k_{#2}}
\def\kk#1#2{k_{#1}\cdot k_{#2}}
\def\G#1#2{G_{B#1#2}}
\def\Gp#1#2{{\dot G_{B#1#2}}}
\def\GF#1#2{G_{F#1#2}}
\def\Dab{{(x_a-x_b)}}
\def\Dsq{{({(x_a-x_b)}^2)}}
\def\lag{( -\partial^2 + V)}
\def\4piTD{{(4\pi T)}^{-{D\over 2}}}
\def\4piT4{{(4\pi T)}^{-2}}
\def\TintmD{{\dps\int_{0}^{\infty}}{dT\over T}\,e^{-m^2T}
    {(4\pi T)}^{-{D\over 2}}}
\def\Tintm4{{\dps\int_{0}^{\infty}}{dT\over T}\,e^{-m^2T}
    {(4\pi T)}^{-2}}
\def\Tintm{{\dps\int_{0}^{\infty}}{dT\over T}\,e^{-m^2T}}
\def\Tint{{\dps\int_{0}^{\infty}}{dT\over T}}
\def\pint{{\dps\int}{dp_i\over {(2\pi)}^d}}
\def\Dx{\dps\int{\cal D}x}
\def\Dy{\dps\int{\cal D}y}
\def\Dpsi{\dps\int{\cal D}\psi}
\def\Tr{{\rm Tr}\,}
\def\tr{{\rm tr}\,}
\def\sumij{\sum_{i<j}}
\def\freeexp{{\rm e}^{-\int_0^Td\tau {1\over 4}\dot x^2}}
\def\arraystretch{2.5}
\def\Ge{\mbox{GeV}}
\def\dA{\partial^2}
\def\DA{\sqsubset\!\!\!\!\sqsupset}
\font\tame = cmmi12 scaled\magstep1
\font\tamt = cmmi12 scaled\magstep2
%
%
\def\bbbr{{\rm I\!R}}
\def\bbbone{{\mathchoice {\rm 1\mskip-4mu l} {\rm 1\mskip-4mu l}
{\rm 1\mskip-4.5mu l} {\rm 1\mskip-5mu l}}}
\def\bbbz{{\mathchoice {\hbox{$\sf\textstyle Z\kern-0.4em Z$}}
{\hbox{$\sf\textstyle Z\kern-0.4em Z$}}
{\hbox{$\sf\scriptstyle Z\kern-0.3em Z$}}
{\hbox{$\sf\scriptscriptstyle Z\kern-0.2em Z$}}}}

\pagestyle{empty}
\renewcommand{\thefootnote}{\fnsymbol{footnote}}
\hskip 9cm {\sl LAPTH--703/98}
\vskip-.1pt
\hskip 9cm {\sl HD-THEP-98/52}
\vskip .9cm
\begin{center}
{\Large\bf 
World Line Path Integrals as a Calculational Tool 
in Quantum Field Theory
\footnote{Talk given by C. S. 
at 6th International Conference on Path Integrals from peV to TeV: 
50 years from
Feynman's Paper, Florence, Italy, 25-29 Aug 1998.} 
}
\vskip1.3cm
{\large 
Michael G. Schmidt}
\\[1.5ex]
{\it
Institut f\"ur Theoretische Physik\\
Universit\"at Heidelberg\\
 Philosophenweg 16\\
D-69120 Heidelberg\\ 
Germany\\
M.G.Schmidt@thphys.uni-heidelberg.de 
\\
}
\vskip1.2cm
\underline{\large Christian Schubert}
\\[1.5ex]
{\it
Laboratoire d'Annecy-le-Vieux
de Physique de Particules\\
Chemin de Bellevue,
BP 110\\
F-74941 Annecy-le-Vieux CEDEX\\
France\\
schubert@lapp.in2p3.fr\\
}
\vskip 1.5cm
\end{center}
\begin{quotation}
\noindent
{\bf Abstract:} We report on the status of the 
string-inspired world line
path integral formalism, a recently developed powerful tool for
the reorganisation of standard perturbative amplitudes in
quantum field theory. The method is
outlined and the present range of its applicability surveyed.
The emphasis is on QED and
QCD photon/gluon amplitudes, with a short discussion of 
axial couplings.
\end{quotation}
\clearpage
\renewcommand{\thefootnote}{\protect\arabic{footnote}}
\pagestyle{plain}

\setcounter{page}{1}
\setcounter{footnote}{0}


%


\section{Historical Remarks}

50 years ago Feynman developed, in the paper which we are celebrating
at this conference \cite{feyn48}, the formulation of quantum
mechanics in terms of path integrals. In the following years
he invented the manifestly relativistic  
framework for perturbative
calculations in second-quantized
quantum field theory which we are still using
today. What is perhaps not so well-known is that, at the time,
Feynman was also experimenting with an alternative
approach to quantum field theory based on his
previous work on the quantum mechanical path integral.
While he did not publish much about this 
first-quantized approach, in
the appendix A of ~\cite{feynman:pr80} he shortly
discusses it 
for the case of scalar quantum electrodynamics,
``for its own interest as an alternative to the formulation of
second quantization''. What he states here is that
the amplitude for a charged scalar particle to move,
under the influence of the external potential $A_{\mu}$,
from point $x$ to $x'$ in Minkowski space
is given by

\bear
\int_0^{\infty}
dT
\int
\!
_{x(0)=x}^{x(T)=x'}
\!
{\cal D}x(\tau)
{\rm e}^{
-{1\over 2}
im^2T}
\exp
\biggl[
-{i\over 2}
\int_0^Td\tau
{({dx_{\mu}\over d\tau})}^2
-i\int_0^T
d\tau
{dx_{\mu}\over d\tau}
A_{\mu}(x(\tau))\nonumber\\
\!\!\!\!\!\!\!\!\!\!\!\!\!\!\!\!
\!\!\!\!\!\!\!\!\!\!\!\!\!\!\!\!
-{i\over 2}
e^2
\int_0^Td\tau
\int_0^Td\tau'
{dx_{\mu}\over d\tau}
{dx_{\nu}\over d\tau'}
\delta_{+}(x(\tau)-x(\tau'))
\biggr]
\qquad
\label{feynform}
\ear
That is,
for a fixed value of the proper-time $T$ 
one can construct the amplitude as a certain
quantum mechanical path integral. This path integral 
is to be performed
over the set of all open
trajectories running from $x$ to $x'$
in the given proper-time.
The action 
consists of the familiar kinetic term, and two interaction terms.
Of those the first one represents the interaction with the external field,
to all orders in the field, while the second one describes an
arbitrary number of virtual photons emitted and re-absorbed along the
trajectory of the particle ($\delta_{+}$ denotes the photon
propagator). This
simple path integral formula
thus corresponds to an infinite number
of Feynman diagrams. Feynman then shows how
to extend it to a path integral formula for the
complete scalar QED S-matrix.
Extensions to spinor QED were also constructed by Feynman
and others ~\cite{refspinorpi}.

While Feynman himself did not make further use of those formulas,
other authors over the years applied them to
a variety of problems
in quantum field theory,
ranging from QED ~\cite{refqed} to
nonabelian gauge theory ~\cite{hajese}, anomalies ~\cite{refanom},
and meson-nucleon theory ~\cite{rossch}.
Still it is fair to say that this approach
never became really popular, nor did any standard way for
calculating this type of path integral emerge.

Renewed interest in the first-quantized approach
was generated in recent years following the work of
Bern and Kosower,
who succeeded in deriving new rules for the construction
of one-loop QCD amplitudes by analyzing the infinite string tension
limits of the corresponding amplitudes in an appropriate
string model ~\cite{berkos}. Those string amplitudes
are represented in terms of the Polyakov path integral,
a first-quantized path integral of the same type as 
Feynman's ``world line'' path integral
above. It is thus not too surprising that
it turned out to be possible ~\cite{strassler}
to rederive those ``Bern-Kosower Rules''
by representing the QED/QCD one-loop effective action
in terms of such path integrals, and evaluating them
in a way analogous to string theory ~\cite{polbook}. 

\section{Photon Scattering in Quantum Electrodynamics}
\renewcommand{\theequation}{2.\arabic{equation}}
\setcounter{equation}{0}

In the case of spinor QED, the 
generalization of eq.(\ref{feynform})
appropriate in the ``stringy'' context is the
following,

\begin{equation}
\Gamma\lbrack A\rbrack = \!-\!2\!\int_0^{\infty}\!
{dT\over T} e^{-m^2T} \!\int \!{\cal D} x {\cal D}\psi\,
{\rm exp}\biggl[\!-\!\!\int_0^T \!\!\!\!\!d\tau
\Bigl({1\over 4}{\dot x}^2 
\!+\! {1\over2}\psi\dot\psi
\!+\! ieA_{\mu}\dot x^{\mu} 
\!-\! ie \psi^{\mu}F_{\mu\nu}\psi^{\nu} \Bigr)\!\biggr]
\label{spinorpi}
\end{equation}

\noindent
Here $\Gamma[A]$ denotes the one-loop (Euclidean)
effective action for the Maxwell field due to an
electron loop. Now we have, in addition to a
path integral over the closed loops
in spacetime $x^{\mu}(\tau ),x^{\mu}(T)=x^{\mu}(0)$,
an additional path integral over the space
of antiperiodic Grassmann functions $\psi^{\mu}(\tau )$
representing the electron spin.
In the ``string--inspired'' approach,
the path integrals over $x$ and $\psi$ 
are evaluated by 
one-dimensional perturbation theory, using the 
Green's functions 
\begin{eqnarray}
\langle x^{\mu}(\tau_1)x^{\nu}(\tau_2)\rangle
   &=& - g^{\mu\nu}G_B(\tau_1,\tau_2)
= - g^{\mu\nu}\biggl[\mid\!\tau_1\!-\!\tau_2\!\mid-
{{(\tau_1\!-\!\tau_2)}^2\over T}\biggr]\nonumber\\
\langle \psi^{\mu}(\tau_1)\psi^{\nu}(\tau_2)\rangle
   &=&{1\over 2}\, g^{\mu\nu} G_F(\tau_1,\tau_2)
= {1\over 2}\, g^{\mu\nu}{\rm sign}
(\tau_1 -\tau_2 )
\label{Green}
\end{eqnarray}
\noindent
(The bosonic Wick contraction is 
actually carried out in the relative coordinate
of the loop, while
the integration over its average
position 
yields energy--momentum conservation.)

One--loop scattering amplitudes are obtained
by specializing the background to 
a finite sum of plane waves of definite
polarization. In the case of scalar QED this
leads to the following 
extremely compact
``Bern-Kosower master
formula'' for the one-loop 
(off-shell, dimensionally regularized)
N-photon amplitude,

\begin{eqnarray}
\Gamma[k_1,\varepsilon_1;\ldots;k_N,\varepsilon_N]
&=&
{(-ie)}^N
{\dps\int_{0}^{\infty}}{dT\over T}
{[4\pi T]}^{-{D\over 2}}
e^{-m^2T}
\prod_{i=1}^N \int_0^T 
d\tau_i
\qquad
\nonumber\\
&&
\!\!\!\!\!\!\!\!\!
\!\!\!\!\!\!\!\!\!
\!\!\!\!\!\!\!\!\!\!\!\!\!\!\!\!\!\!\!\!\!\!\!\!\!
\!\!\!\!\!\!\!\!\!\!\!\!\!\!\!\!\!\!\!\!\!\!\!\!\!
\times
\exp\biggl\lbrace \sum_{i,j=1}^N 
\bigl\lbrack \half G_{Bij} k_i\cdot k_j
-i\dot G_{Bij}\varepsilon_i\cdot k_j
+\half\ddot G_{Bij}\varepsilon_i\cdot\varepsilon_j
\bigr\rbrack\biggr\rbrace
\mid_{\rm multi-linear}
\label{scalarqedmaster}
\end{eqnarray}
\no
($G_B(\tau_1,\tau_2) \equiv G_{B12}$ etc.).
Here it is understood that only the terms linear
in all the $\varepsilon_1,\ldots,\varepsilon_N$
have to be taken. $D$ denotes the spacetime dimension.
Besides the Green's function $G_B$ also its first and
second derivatives appear,
$\dot G_B(\tau_1,\tau_2) = {\rm sign}(\tau_1 - \tau_2)
- 2 {{(\tau_1 - \tau_2)}\over T},
\ddot G_B(\tau_1,\tau_2)
= 2 {\delta}(\tau_1 - \tau_2)
- {2\over T}
$.
Dots generally denote a
derivative acting on the first variable.
For the fermion QED case an analogous formula can be
written using a superfield formalism \cite{ss3,rss}.
Alternatively the additional 
terms from the Grassmann path integration
can also be generated
by performing a certain partial integration
algorithm on the above expression, and then
applying a simple ``substitution rule'' on the
result \cite{berkos}. 
For example, the following representation is obtained
after partial integration for the one-loop 
4-photon amplitude in scalar QED ~\cite{Nphoton},

\bear
\Gamma
[\lbrace k_i,\varepsilon_i\rbrace ]
&=&
e^4
{\dps\int_{0}^{\infty}}{dT\over T}
{[4\pi T]}^{-{D\over 2}}
e^{-m^2T}
\prod_{i=1}^4 \int_0^T 
d\tau_i\,
\bigl(
Q_4^4+Q_4^3+Q_4^2-Q_4^{22}
\bigr)
\e^{\half
G_{Bij} k_i\cdot k_j
}
\non\\
Q_4^4 &=& 
\dot G_{B12}
\dot G_{B23}
\dot G_{B34}
\dot G_{B41}
Z_4(1234)
+ 2 \,\, {\rm permutations}
\non\\
Q_4^3 &=&
\dot G_{B12}
\dot G_{B23}
\dot G_{B31}
Z_3(123)
\dot G_{B4i}
\varepsilon_4\cdot k_i
+ 3 \,\, {\rm perm.}
\non\\
Q_4^2 &=&
\dot G_{B12}\dot G_{B21}
Z_2(12)
\biggl\lbrace
\dot G_{B3i}
\varepsilon_3\cdot k_i
\dot G_{B4j}
\varepsilon_4\cdot k_j
\non\\
&&
+\half
\dot G_{B34}
\varepsilon_3\cdot\varepsilon_4
\Bigl[
\dot G_{B3i}
k_3\cdot k_i
-
\dot G_{B4i}
k_4\cdot k_i
\Bigr]
\biggr\rbrace
+ \, 5 \,\, {\rm perm.}
\non\\
Q_4^{22} &=&
\dot G_{B12}\dot G_{B21}
Z_2(12)
\dot G_{B34}\dot G_{B43}
Z_2(34)
+ 2 \,\, {\rm perm.}
\non\\
\label{4photon}
\ear\no
Here summation over dummy indices from $1,\ldots,4$
is understood, and

\bear
Z_2(ij)&\equiv&
\varepsilon_i\cdot k_j
\varepsilon_j\cdot k_i
-\varepsilon_i\cdot\varepsilon_j
k_i\cdot k_j
\non\\
Z_n(i_1i_2\ldots i_n)&\equiv&
{\rm tr}
\prod_{j=1}^n 
\Bigl[
k_{i_j}\otimes \varepsilon_{i_j}
- \varepsilon_{i_j}\otimes k_{i_j}
\Bigr]
\quad (n\geq 3)
\non
\ear\no
This formula has
all properties which one could possibly demand
from an integral representation for this amplitude,
namely
i) it provides a maximal gauge invariant decomposition,
ii) manifest term-by-term UV finiteness,
iii) permutation symmetry,
iv) it represents the complete amplitude rather than
a single Feynman diagram. 
The full set of Bern-Kosower rules allows one to use it
to construct, by mere
pattern matching,
not only its spinor QED equivalent, but also the
corresponding gluonic amplitudes.

\section{QED in a Constant Field}
\renewcommand{\theequation}{3.\arabic{equation}}
\setcounter{equation}{0}

The presence of an additional constant external field
$F_{\mu\nu}$ can be shown to
change the master formula
eq.(\ref{scalarqedmaster})
to
\cite{rss,shaisultanov}

\begin{eqnarray}
\Gamma[k_1,\varepsilon_1;\ldots;k_N,\varepsilon_N]
&=&
{(-ie)}^N
{\dps\int_{0}^{\infty}}{dT\over T}
{[4\pi T]}^{-{D\over 2}}
e^{-m^2T}
{\rm det}^{-{1\over 2}}
\biggl[{{\rm sin}(eFT)\over eFT}\biggr]
\nonumber\\
&&
\!\!\!\!\!\!\!\!\!\!\!\!\!\!\!\!\!\!\!\!\!\!\!\!\!\!\!\!\!
\!\!\!\!\!\!\!\!\!\!\!\!\!\!\!\!\!\!\!\!\!\!\!\!\!\!\!\!\!
\!\!\!\!\!
\times
\prod_{i=1}^N \int_0^T 
d\tau_i
\exp\biggl\lbrace\sum_{i,j=1}^N 
\bigl\lbrack \half k_i\cdot {\cal G}_{Bij}\cdot  k_j
-i\varepsilon_i\cdot\dot{\cal G}_{Bij}\cdot k_j
+\half
\varepsilon_i\cdot\ddot {\cal G}_{Bij}\cdot\varepsilon_j
\bigr\rbrack\biggr\rbrace
\mid_{\rm multi-linear}
\nonumber\\
\label{scalarqedmasterF}
\end{eqnarray}
\no
where
$z=eFT$, and

\bear
{{\cal G}_B}(\tau_1,\tau_2) &=&
{T\over 2z^2}
\biggl({z\over{{\rm sin}(z)}}
\,{\rm e}^{-iz\dot G_{B12}}
+ iz\dot G_{B12} - 1\biggr)
\non\\
{\cal G}_{F}(\tau_1,\tau_2) &=&
G_{F12} {{\rm e}^{-iz\dot G_{B12}}\over {\rm cos}(z)}
\label{defcalGBGF}
\ear\no
For example,
using this formula with $N=2$ it takes only a few
lines to calculate the QED vacuum polarization tensor
in a general constant field, 
a calculation which in field theory is
already substantial.
The $N=3$ case was used in ~\cite{adlsch}
for a recalculation of the photon splitting amplitude
in a magnetic field, and also showed a significant
gain in efficiency.

Much less explored are presently the QED amplitudes involving external
scalars ~\cite{dashsu} or fermions ~\cite{mckreb}, for which
no equally elegant formulation has been found so far as in the
photon case.

\section{Other Field Theories}
\renewcommand{\theequation}{4.\arabic{equation}}
\setcounter{equation}{0}

Eq.(\ref{spinorpi}) generalizes to the case of an external
non-abelian gauge field simply by the addition of
a colour trace, and the path-ordering operator.
The accomodation of internal gluons is also possible
but more involved; it requires the introduction of
auxiliary degrees of freedom in the loop whose
contributions have to be projected out
~\cite{strassler,rss}. 

While worldline representations for gauge couplings have
been discussed for decades, 
only following the development of the ``string-inspired''
formalism systematic searches for
generalizations to other field theories 
were undertaken. Appropriate path integrals 
are now available for the fermion loop coupled to external
(pseudo-)scalars ~\cite{mnss1,dhogag1}, axial-vectors
~\cite{mnss2,dhogag2}, and antisymmetric tensors
~\cite{dhogag2}. Those constructions were based on a dimensional
reduction procedure from six-dimensional gauge theory.
A particularly simple and direct construction exists for
the case where only a vector field $A$ and axial-vector field
$A_5$ are present ~\cite{mcksch}. It expresses the effective
action $\Gamma[A,A_5]$ in terms of the same path integral
formula as in eq.(\ref{spinorpi}), with the worldline
Lagrangian replaced by

\bear
L(\tau) &\longrightarrow&
L(\tau) 
-2i\hat\gamma_5\dot x^{\mu}\psi_{\mu}\psi_{\nu}A_5^{\nu}
+i\hat\gamma_5\partial_{\mu}A^{\mu}_5
+(D-2)A_5^2
\label{LAA5}
\ear\no
Here the operator $\hat\gamma_5$ has the effect of
flipping the boundary conditions for the Grassmann
path integral from antiperiodic to periodic.
In the presence of pseudo-scalars or axial-vectors
the evaluation of fermion loops is expected to give rise
to $\varepsilon$ -- tensors. In the worldline formalism
those are produced by the Grassmann zero mode
integral which one has for periodic boundary conditions.
The above worldline Lagrangian turned out to be
very suitable
to the calculation of the vector -- axial vector amplitude
in a constant external field ~\cite{ioasch}.

\noindent
For an application of the pseudoscalar path integral
to axion decay see ~\cite{haasch}.
Steps towards an extension to curved backgrounds
were undertaken in ~\cite{dilmck}. 

\section{Multiloop Extension}
\renewcommand{\theequation}{5.\arabic{equation}}
\setcounter{equation}{0}

Since the one-loop parameter integral representations
obtained in the worldline formalism are generally
valid off-shell, they can be used to construct higher -- loop
amplitudes by sewing. A more elegant route to
multi-loop extension is provided by the introduction
of worldline multiloop Green's functions 
~\cite{ss2}.
Those are effective worldline propagators taking the
effect of propagator insertions into the one-loop graph
into account. As shown in ~\cite{rolsat}
they are the leading-order coefficients 
of the corresponding string-theoretic worldsheet Green's
functions in the ${1\over{\alpha'}}$ -- expansion.

In either case one arrives at integral representations for
multiloop amplitudes in $\phi^3$ -- theory
~\cite{ss2,satsch}, QED ~\cite{ss3,dashsu}
or QCD ~\cite{satoqcd}
that are of a similar form as
our one-loop formulas above.
However here we find the
additional interesting feature that a single worldline
parameter integral may contain contributions from many
Feynman diagrams of different topologies. 
In ~\cite{ss3} the usefulness
of this property was demonstrated for the case of the
2-loop scalar and spinor QED $\beta$ -- functions.
The concept of multiloop worldline propagators can be
generalized to the constant external field case
as in the one-loop case. 
This was used for recalculations
of the 2-loop scalar and spinor QED
Euler-Heisenberg Lagrangians
~\cite{rss,frsskoesch}. 

\section{Conclusions}

To summarize, by now there is sufficient practical
experience indicating that the worldline path
integral formalism is an excellent tool for the
calculation of the QED photon S-matrix in general,
and definitely superior to standard field theory
for problems involving constant external fields.
While here we have concentrated on amplitude
calculations, similar improvements on field theory
were found also in calculations of the effective
action itself in the inverse mass expansion
~\cite{refhde}.
By now a variety of extensions to other types of amplitudes
exist, though those
have not been sufficiently tested yet to
allow for general statements on their efficiency
as a calculational tool.


\begin{thebibliography}{99}

\bibitem{feyn48} R.P. Feynman, {\it Rev. Mod. Phys.} {\bf 20},
367 (1948).

\bibitem{feynman:pr80} R.P. Feynman, {\it Phys. Rev.} 
{\bf 80}, 440 (1950).

\bibitem{refspinorpi} 
R.P. Feynman, {\it Phys. Rev.} {\bf 84}, 108 (1951); 
E.S. Fradkin, {\it Nucl. Phys.} {\bf 76}, 588 (1966);
F.A. Berezin and M.S. Marinov, {\it Ann. Phys.} {\bf 104}, 336 (1977);
F. Bordi and R. Casalbuoni,
{\it Phys. Lett.} {\bf B93}, 308 (1980).

\bibitem{refqed}
H.M. Fried, {\it Functional Methods and Models in Quantum Field 
Theory}, MIT Press, Cambridge (1972);
A. Barducci, F. Bordi, and R. Casalbuoni,
{\it Nuov. Cim.} {\bf B 64}, 287 (1981).

\bibitem{hajese}
B. Halpern, A. Jevicki, and P. Senjanovic, {\it Phys. Rev.} 
{\bf D16}, 2476 (1977).

\bibitem{refanom}
L. Alvarez-Gaum\'e, {\it Comm. Math. Phys.} {\bf 90}, 161
(1983);
F. Bastianelli and P. van Nieuwenhuizen, {\it Nucl. Phys.}
{\bf B389}, 53 (1993). 

\bibitem{rossch}
R. Rosenfelder and A.W. Schreiber, 
{\it Phys. Rev.} {\bf D53}, 3337 (1996)
(nucl-th/9504002). 

\bibitem{berkos}
Z. Bern and D. A. Kosower,
{\it Nucl. Phys.} {\bf B379}, 451 (1992).

\bibitem{strassler}
M. J. Strassler, {\it Nucl. Phys.} {\bf B385}, 145 (1992).

\bibitem{polbook}
A. M. Polyakov, {\sl Gauge Fields and
Strings}, Haarwood 1987.

\bibitem{ss3}
M.G. Schmidt and C. Schubert, {\it Phys. Rev.} {\bf D53}, 2150
(1996) (hep-th/9410100).

\bibitem{rss}
M. Reuter, M.G. Schmidt, and C. Schubert,
{\it Ann. Phys.} {\bf 259}, 313 (1997) (hep-th/9610191).

\bibitem{Nphoton}
C. Schubert, {\it Europ. J. Phys.} {\bf C}, in print
(hep-th/9710067).  

\bibitem{shaisultanov}
R. Shaisultanov, Phys. Lett. {\bf B 378}, 354 
(1996) (hep-th/9512142).

\bibitem{adlsch}
S. L. Adler and C. Schubert, {\it Phys. Rev. Lett.}
{\bf 77}, 1695 (1996) (hep-th/9605035).

\bibitem{dashsu}
K. Daikouji, M. Shino, and Y. Sumino, 
{\it Phys. Rev.} {\bf D53}, 4598 (1996) (hep-ph/9508377). 

\bibitem{mckreb}
D.G.C. McKeon and A. Rebhan, {\it Phys. Rev.} {\bf D48}, 2891 (1993).

\bibitem{mnss1}
M. Mondrag\'on, L. Nellen, M.G. Schmidt, and C. Schubert,
{\it Phys. Lett.} {\bf B351}, 200 (1995) (hep-th/9502125).

\bibitem{dhogag1}
E. D' Hoker and D. G. Gagn\'e, 
{\it Nucl. Phys.} {\bf B467}, 272 (1996) (hep-th/9508131).

\bibitem{mnss2}
M. Mondrag\'on, L. Nellen, M.G. Schmidt, and C. Schubert,
{\it Phys. Lett.} {\bf B366}, 212 (1996) (hep-th/9510036).

\bibitem{dhogag2}
E. D' Hoker and D. G. Gagn\'e, 
{\it Nucl. Phys.} {\bf B467}, 297 (1996) (hep-th/9512080).

\bibitem{mcksch}
D.G.C. McKeon and C. Schubert, 
{\it Phys. Lett.} {\bf B}, in print
(hep-th/9807072). 

\bibitem{ioasch}
A.N. Ioannisian and C. Schubert, in preparation;
C. Schubert, LAPTH-687-98 (hep-ph/9807288). 

\bibitem{haasch}
M. Haack and M.G. Schmidt, 
Europ. J. Phys. {\bf C}, in print (hep-th/9806138). 

\bibitem{dilmck}
F.A. Dilkes and D.G.C. McKeon, 
Phys. Rev. {\bf D 53}, 4388 (1996) 
(hep-th/9509005). 

\bibitem{ss2}
M. G. Schmidt and C. Schubert, {\it Phys. Lett.} {\bf B331}, 69
(1994) (hep-th/9403158).

\bibitem{rolsat}
K. Roland and H.-T. Sato, {\it Nucl. Phys.} {\bf B 480}, 99
(1996) (hep-th/9604152).   

\bibitem{satsch}
H.-T. Sato and M.G. Schmidt, {\it  Nucl. Phys.} {\bf B 524}, 
742 (1998) (hep-th/9802127).  

\bibitem{satoqcd}
H.-T. Sato, {\it Nucl. Phys.} {\bf B 491}, 477 (1997) (hep-th/9610064). 

\bibitem{frsskoesch}
D. Fliegner, M. Reuter, M.G. Schmidt, and C. Schubert,
{\it Theor. Math. Phys.} {\bf 113}, 1442 (1997) (hep-th/9704194);
B. K\"ors and M.G. Schmidt,
Europ. J. Phys. {\bf C}, in print (hep-th/9803144).  

\bibitem{refhde}
M. G. Schmidt and C. Schubert, {\it Phys. Lett.} {\bf B318}, 438
(1993) (hep-th/9309055);
D. Fliegner, M.G. Schmidt, and C. Schubert,
{\it Z. Phys.} {\bf C 64}, 111 (1994) (hep-ph/9401221);
and with P. Haberl, {\it  Ann. Phys.} {\bf 264}, 51 (1998)
(hep-th/9707189);
D. Cangemi, E. D'Hoker, and G. Dunne, {\it Phys. Rev.} 
{\bf D 51}, 2513 (1995) (hep-th/9409113);
V.P. Gusynin and I.A. Shovkovy, 
UCTP-106-98 (hep-th/9804143). 

\end{thebibliography}
\end{document}